\documentclass{article}
\usepackage[utf8]{inputenc}
\usepackage{ijcai25}

\usepackage{times}
\usepackage{soul}
\usepackage{url}
\usepackage[hidelinks]{hyperref}
\usepackage[utf8]{inputenc}
\usepackage[small]{caption}
\usepackage{graphicx}
\usepackage{amsmath}
\usepackage{amsthm}
\usepackage{booktabs}
\usepackage{algorithm}
\usepackage{algorithmic}

\usepackage[switch]{lineno}

\usepackage{tabularx}

\usepackage{caption}
\title{Efficient Quantum Approximate $k$NN Algorithm via Granular-Ball Computing}

\author{
    Author Name
    \affiliations
    Affiliation
    \emails
    email@example.com
}

\author{
Shuyin Xia$^{1,3}$ \and
Xiaojiang Tian$^2$ \and
Suzhen Yuan$^{2,3*}$ \and
Jeremiah D. Deng$^4$\\
\affiliations{
  $^1$School of Artificial Intelligence, Chongqing University of Posts and Telecommunications, Chongqing 400065, China \\
  $^2$School of Electronic Science and Engineering, Chongqing University of Posts and Telecommunications, Chongqing 400065, China\\
  $^3$Key Laboratory of Big Data Intelligent Computing, Chongqing University of Posts and\\ Telecommunications, Chongqing, 400065, China\\
  $^4$School of Computing, University of Otago, Dunedin 9054, New Zealand \\
}
\emails{
  xiasy@cqupt.edu.cn,
  s230401037@stu.cqupt.edu.cn,
  yuansuzhen@cqupt.edu.cn,
  jeremiah.deng@otago.ac.nz
}
}

\begin{document}

\maketitle
\begin{abstract}
High time complexity is one of the biggest challenges faced by $k$-Nearest Neighbors ($k$NN). Although current classical and quantum $k$NN algorithms have made some improvements, they still have a speed bottleneck when facing large amounts of data. To address this issue, we propose an innovative algorithm called Granular-Ball based Quantum $k$NN(GB-Q$k$NN). This approach achieves higher efficiency by first employing granular-balls, which reduces the data size needed to processed. The search process is then accelerated by adopting a Hierarchical Navigable Small World (HNSW) method. Moreover, we optimize the time-consuming steps, such as distance calculation, of the HNSW via quantization, further reducing the time complexity of the construct and search process. By combining the use of granular-balls and quantization of the HNSW method, our approach manages to take advantage of these treatments and significantly reduces the time complexity of the $k$NN-like algorithms, as revealed by a comprehensive complexity analysis.
\end{abstract}

\section{Introduction}

Machine learning, a fundamental technology of artificial intelligence, facilitates prediction and decision-making by extracting patterns from large data sets. Within the supervised learning domain, the $k$NN algorithm \cite{guo-knn-model} has been widely adopted for classification and regression due to its conceptual simplicity and practical effectiveness. However, the rapid expansion of the data scale in the current era of information has exposed significant challenges associated with the time complexity of the classical $k$NN algorithm, particularly when applied to large-scale data sets. This challenge has emerged as a critical bottleneck that limits the algorithm's scalability and broader applicability.

This paper seeks to address these limitations by improving algorithmic efficiency while maintaining the accuracy of nearest neighbor search, utilizing two pivotal approaches: granular-ball theory \cite{xia-granular-ball} and the quantization of HNSW \cite{malkov-hnsw}. As an advanced data modeling technique, granular-ball theory effectively reduces data volume and alleviates the computational burden of large-scale data sets through granularity division and aggregation processes. In addition, this paper employs quantum computing by encoding information in qubits and harnessing their unique properties of superposition and entanglement.

Specifically, for the GB-Q$k$NN algorithm proposed in our paper, the granular-ball theory is applied to reduce the data scale and represent the original data set using the granular-ball, thus significantly decreasing the graph construction scale. Subsequently, a quantum algorithm is employed to optimize the HNSW method. Once the graph is constructed, the quantum computation performs layer-by-layer $k$NN searches on the test data set, completing the classification tasks efficiently. This approach achieves a graph construction complexity of $O(M\log{M})$, where $N$ is the size of the data set, and $M (M \ll N)$  is the number of granular-ball. In addition, it can reduce the search complexity to $O(\log{M})$. Compared to existing classical and quantum $k$NN algorithms, the proposed approach demonstrates superior efficiency and accuracy, making it a compelling solution for large-scale data sets.
Our contributions are as follows.

\noindent\hspace*{2em}\textbullet\ This work integrates the granular-ball theory with the HNSW method, effectively reducing the data size and significantly accelerating the search process for $k$NN.\\
\noindent\hspace*{2em}\textbullet\ We extend the HNSW method by introducing a quantum-enhanced approach, taking advantage of the inherent parallelism of quantum computing to accelerate distance computations and significantly improve graph construction efficiency. \\
\noindent\hspace*{2em}\textbullet\ A comprehensive time complexity analysis reveals that the proposed algorithm achieves a lower time complexity than existing classical and quantum $k$NN algorithms.

\section{Related Work}
The classical $k$NN algorithm, first introduced by Thomas Cover in 1967, has experienced substantial development due to its intuitive design and high accuracy. However, exponential growth in data size has rendered its time complexity a critical bottleneck, limiting its practical applicability. To solve this problem, various approximate $k$NN algorithms have been proposed. The HNSW method, introduced in 2016, employs a graph-based hierarchical structure with small-world properties. With a construction time complexity of $O(N\log{N})$, HNSW achieves a significant reduction in the search time complexity to $O(\log{N})$. In 2021, Fu advanced this line of research by proposing the NSSG \cite{fu-satellite-graph} (Navigable Satellite System Graphs), which builds on the SSG (Satellite System Graphs) through an optimized pruning strategy. This approach achieves a construction time complexity of $O(N + N^{1.14})$ while maintaining a search time complexity of $O(\log{N})$. Currently, quantum machine learning is rapidly advancing, holds significant promise, and has driven the development of advanced algorithms such as quantum support vector machines \cite{rebentrost-qsvm}, quantum $k$NN \cite{basheer-quantum-knn}, and quantum neural networks \cite{altaisky-qnn}. In particular, for quantum $k$NN, Manuela C. G. Laing proposed a preliminary quantum $k$NN framework leveraging Grover’s algorithm in 2017, attaining a time complexity of $O(\sqrt{N})$. Building on this, in 2022, a quantum $k$ NN algorithm based on the Hamming distance \cite{li-hamming-knn} was introduced, achieving a time complexity of $O(\sqrt{N}\log^{2}{N})$ and marking a notable improvement. In 2023, a quantum $k$NN algorithm employing polar distance \cite{feng-polar-distance} further reduced the time complexity to $O(\sqrt{kN})$.

The structure of this paper is organized as follows. Section 3 introduces the foundational concepts of granular-ball, quantum computing, and the HNSW method. Section 4 provides a detailed description of the proposed algorithm and its implementation process. Section 5 presents a comprehensive analysis of the algorithm's time complexity and a performance comparison with existing classical approximate $k$NN and quantum $k$NN algorithms. Finally, Section 6 concludes the paper by summarizing key contributions and highlighting potential directions for future research in related areas.

\section{Background}
This section provides an overview of the fundamentals for the proposed algorithm, including the concept of granular-balls, 
quantum computing basics, and the HNSW method.

\subsection{Granular-Balls}  
Granular-ball computation is an efficient and robust granular computing method proposed in \cite{xia-granular-ball} based on granular cognitive computing, which can also reduce attributes. The core idea is to use granular-balls to cover and represent the sample space. For a granular-ball $\mathcal{G} = \{ \mathbf{x}_i, i = 1,\cdots,n\} $, where $\mathbf{x}_i$ represents a sample point within the granular-ball, and $n$ represents the number of sample points in the granular-ball, the center $C$ and radius $R$ of the granular-ball can be described as:
\begin{equation}
	\mathbf{C} = \frac{1}{n}\sum\limits_{i = 1}^n {{\mathbf{x}_i}} ,	
\end{equation}
\begin{equation}
	R = \frac{1}{n}\sum\limits_{i = 1}^n \| \mathbf{x}_i -\mathbf{C} \| .
\end{equation}

\subsection{Quantum Algorithms}  

Now we introduce some quantum computing circuits used in our algorithm.

\subsubsection{QRAM}
QRAM (Quantum Random Access Machine) ~\cite{giovannetti-qram} leverages the properties of quantum superposition and entanglement to implement a quantum memory capable of random access. It stores addresses and data in the form of quantum entanglement. The specific data encoding process is as follows:
\begin{equation}
	\label{formula 1}
	\sum\limits_{j} \psi_j{\vert j \rangle}_a \xrightarrow{QRAM}\sum\psi_j {\vert j \rangle}_a {\vert D_j \rangle}_a
\end{equation}
where ${\left| j \right\rangle _a}$ represents the $j$-th memory unit. After being stored in QRAM, the data are encoded in the data register. ${\left| {{D_j}} \right\rangle _a}$ represents the value of the $j$-th data D, and ${\psi _j}$ denotes the probability amplitude of each state.

\subsubsection{Angle encoding}
Angle encoding \cite{farhi-fixed-architectures} utilizes $RX(\theta )$, $RY(\theta )$, and $RZ(\theta )$ to rotate the angles in the Bloch sphere to encode classical information. For quantum algorithms that do not require arithmetic operations, the three above-mentioned rotation gates transform the data into the relative phase angles of the quantum bits, effectively mapping the data to a point on the Bloch sphere. The specific implementation of the controlled RY (CRY) gate 
$\mathrm{CRY}(\cdot)$ is given by
\begin{equation}
    \label{angle encoding}
    \mathrm{CRY}\left| a \right\rangle \left| 0 \right\rangle  \to \left| {00} \right\rangle  + \left| 1 \right\rangle \left[ {\cos (\theta )\left| 0 \right\rangle  + \sin (\theta )\left| 1 \right\rangle } \right],
\end{equation}
where $\vert{a}\rangle$ is the control bit of the CRY gate and $\theta$ is the rotation angle.

\subsubsection{Swap test}
\begin{figure}[!t]
	\centering	\includegraphics[width=0.5\linewidth]{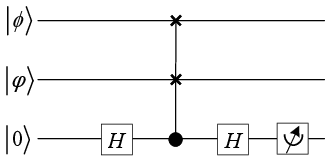}
	\caption{\label{fig_3} Quantum circuit for swap test.}
\end{figure}

Swap test \cite{bennett-teleportation} calculates the similarity between quantum states $\left| \phi  \right\rangle $ and $\left| \varphi  \right\rangle $ by determining the similarity, that is, the squared modulus of the inner product between these states. The quantum circuit of the Swap test is shown in Fig. \ref{fig_3}. The output of this quantum circuit is:
\begin{equation}
    \label{swap test}
	{\vert 0 \rangle}{\vert {\phi} \rangle}{\vert {\varphi} \rangle}\rightarrow\frac{1}{2}{\vert 0 \rangle}({\vert {\phi} \rangle}{\vert {\varphi} \rangle}+{\vert {\varphi} \rangle}{\vert {\phi} \rangle})+\frac{1}{2}{\vert 1 \rangle}({\vert {\phi} \rangle}{\vert {\varphi} \rangle}-{\vert {\varphi} \rangle}{\vert {\phi} \rangle})
\end{equation}
When we measure the last qubit of the output, the following probability can be obtained:
\begin{equation}
	p(0)=\frac{1}{2}+\frac{1}{2}\lvert{\langle {\varphi} \vert {\phi} \rangle}\rvert^2
\end{equation}
\begin{equation}
	p(1)=\frac{1}{2}-\frac{1}{2}\lvert{\langle {\varphi} \vert {\phi} \rangle}\rvert^2
\end{equation}

We can obtain the similarity of two quantum states through the above three quantum modules. Algorithm 1 gives the pseudocode for convenience in the following use. 
\begin{algorithm}[t]
    \footnotesize
	\caption{Similarity Calculation}
	\label{alg:swaptest}
	\textbf{Input}: Data point set $S$, single point $B$\\
	\textbf{Output}: Similarity set $H$
	\begin{algorithmic}[0]
        \STATE $M\leftarrow |B|, N\leftarrow 2^{\lceil\log_2|S|\rceil}$\;
		\STATE Through QRAM, angle encoding map  $S, B$ onto angles: $\sum\limits_{i=0}^N  \sum \limits_{j=0}^M \vert i\rangle  \vert j \rangle \vert{0}\rangle \vert{0} \rangle \rightarrow \sum\limits_{i=0} ^N\sum\limits_{j=0}^M \vert i\rangle   \vert j \rangle \vert{\phi}\rangle \vert{\varphi}\rangle$\;
		\STATE Through Swap test calculate similarity: 
		$\sum\limits_{i=0} ^N\sum\limits_{j=0}^M \vert i\rangle   \vert j \rangle \vert{\phi}\rangle \vert{\varphi}\rangle \vert 0 \rangle \rightarrow \sum\limits_{i=0} ^N\sum\limits_{j=0}^M \vert i\rangle   \vert j \rangle \frac{1}{2}((\vert{\phi}\rangle\vert{\varphi}\rangle+\vert{\varphi}\rangle\vert{\phi}\rangle)\vert{0}\rangle+(\vert{\phi}\rangle\vert{\varphi}\rangle-\vert{\varphi}\rangle\vert{\phi}\rangle)\vert{1}\rangle)$\;
	\end{algorithmic}
\end{algorithm}
\subsubsection{Quantum Comparison Circuit}
A quantum comparator can be used to compare two values. 

\begin{figure}[!t]
	\centering	\includegraphics[width=0.7\linewidth]{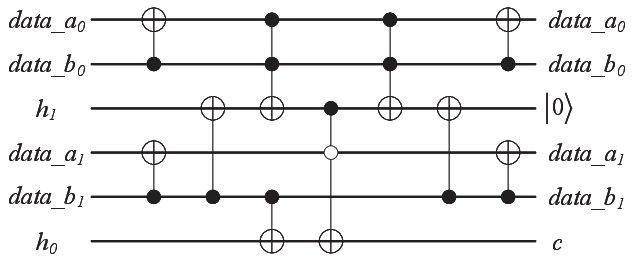}
	\caption{\label{fig_4} Quantum circuit for comparison.}
\end{figure}

The unitary evolution of this quantum state after passing through the quantum comparison circuit is shown in Eq. \ref{qcmp}:
\begin{equation}
    \label{qcmp}
	{U_{CMP}}\left| a \right\rangle \left| b \right\rangle \left| 0 \right\rangle \left| 0 \right\rangle  = \left| a \right\rangle \left| b \right\rangle \left| 0 \right\rangle \left| c \right\rangle 
\end{equation}
In the above equation $\left| a \right\rangle \left| b \right\rangle $ are two numbers involved in the comparison, the $\left| 0 \right\rangle $ represent the the auxiliary bit with input state 0, $\left| c \right\rangle $ stores information about the comparison results, when $c = 0$, $a \ge b$; and when $c = 1$, $a < b$. Figure \ref{fig_4} shows the quantum circuit of the two-bit quantum comparator. 
 
Finally, the comparison result is obtained by the value of $c$ at the end of the output, the details are given in Algorithm 2:
\begin{algorithm}[!b]
    \footnotesize
	\caption{Quantum Compare}
	\label{alg:QuantumCompare}
	\textbf{Input}: Data values $a$ and $b$\\
	\textbf{Output}: Result of $\min(a,b)$
	\begin{algorithmic}[0]
        \STATE $M\leftarrow |b|, N\leftarrow 2^{\lceil\log_2|a|\rceil}$\;
		\STATE Through QRAM, angle encoding map $a, b$ onto angles: $\sum\limits_{i=0}^N  \sum \limits_{j=0}^M \vert i\rangle  \vert j \rangle \vert{0}\rangle \vert{0} \rangle \rightarrow \sum\limits_{i=0} ^N\sum\limits_{j=0}^M \vert i\rangle   \vert j \rangle \vert{\phi}\rangle \vert{\varphi}\rangle$\;
		\STATE QuantumCompare: 
		${U_{CMP}}\left| a \right\rangle \left| b \right\rangle \left| 0 \right\rangle \left| 0 \right\rangle  = \left| a \right\rangle \left| b \right\rangle \left| 0 \right\rangle \left| c \right\rangle $;
		\STATE {Return } $c$ == 0 \, ? \, $b$ : $a$;
	\end{algorithmic}
\end{algorithm}

\subsection{HNSW}  

HNSW is an approximate $k$NN search algorithm inspired by small-world networks, which utilizes a hierarchical graph structure to accelerate the search process. HNSW achieves high query efficiency and scalability by combining local connections with long-range shortcuts, outperforming traditional planar or linear graph representations.

The two main processes of HNSW are construction and search. During construction, nodes are assigned to layers based on a probability distribution and connect to nearest neighbors using a greedy strategy. During the search, the algorithm starts at the top layer, iteratively selects the nearest neighbors, and refines the results at the lower layers to identify the $k$ nearest neighbors. Priority queues optimize neighbor selection, ensuring efficient traversal. These points establish connections or conduct search operations with their neighbors, progressing from the top layer to the bottom.

\section{Methodology}
Before introducing our proposed algorithm, we list the notations used in presenting the algorithm in Table \ref{tab:notations}.
\begin{table}[!t]
    
	\centering
	\LARGE
	\resizebox{\columnwidth}{!}
	{
		\begin{tabularx}{\textwidth}{>{\centering\arraybackslash}p{2cm}|p{15cm}}
			\hline
			\textbf{Notations} & \textbf{Descriptions} \\ \hline
			$k$             & The value $k$ in $k$NN \\ \hline
			$\mathcal{D}$             & Original data set \\ \hline
			$\mathcal{G}$             & Granular-ball data set \\ \hline
			$T$             & Granular-ball purity threshold \\ \hline
			$M$             & The size of Granular-ball data set  \\ \hline
			$L$ & Highest level at which a data point is inserted\\ \hline
			$d$             & Data dimension \\ \hline
			$S$             & Address qubits \\ \hline
			$t_a$             & Data qubits \\ \hline
            $ang$             & angle qubit \\ \hline
			$S_i$           & Set of data points corresponding to each layer graph \\ \hline
			$J_i$           & Adjacency matrix corresponding to each layer graph \\ \hline
			$S_{ci}$ & Set of nodes participating in similarity computation at each layer\\ \hline
			$q$             & Maximum number of binary bits for similarity data \\ \hline
			$d_i$  & Test data points during the search \\ \hline
			$d_{ci}$  & Nearest neighbor of the current layer in the search process \\ \hline
			$d_{bi}$  & Highest priority value in the current layer priority queue \\ \hline
			$m$  & Maximum number of neighbors of a node \\ \hline
            $C$  & The point closest to the insertion point at each layer \\ \hline
            $N$  & The size of original data set \\ \hline
		\end{tabularx}
	}
        \centering\caption{\label{tab:notations}Used notations}
\end{table}

\subsection{Algorithm Flow}

\begin{figure}[!t]
	\centering	\includegraphics[width=1\linewidth]{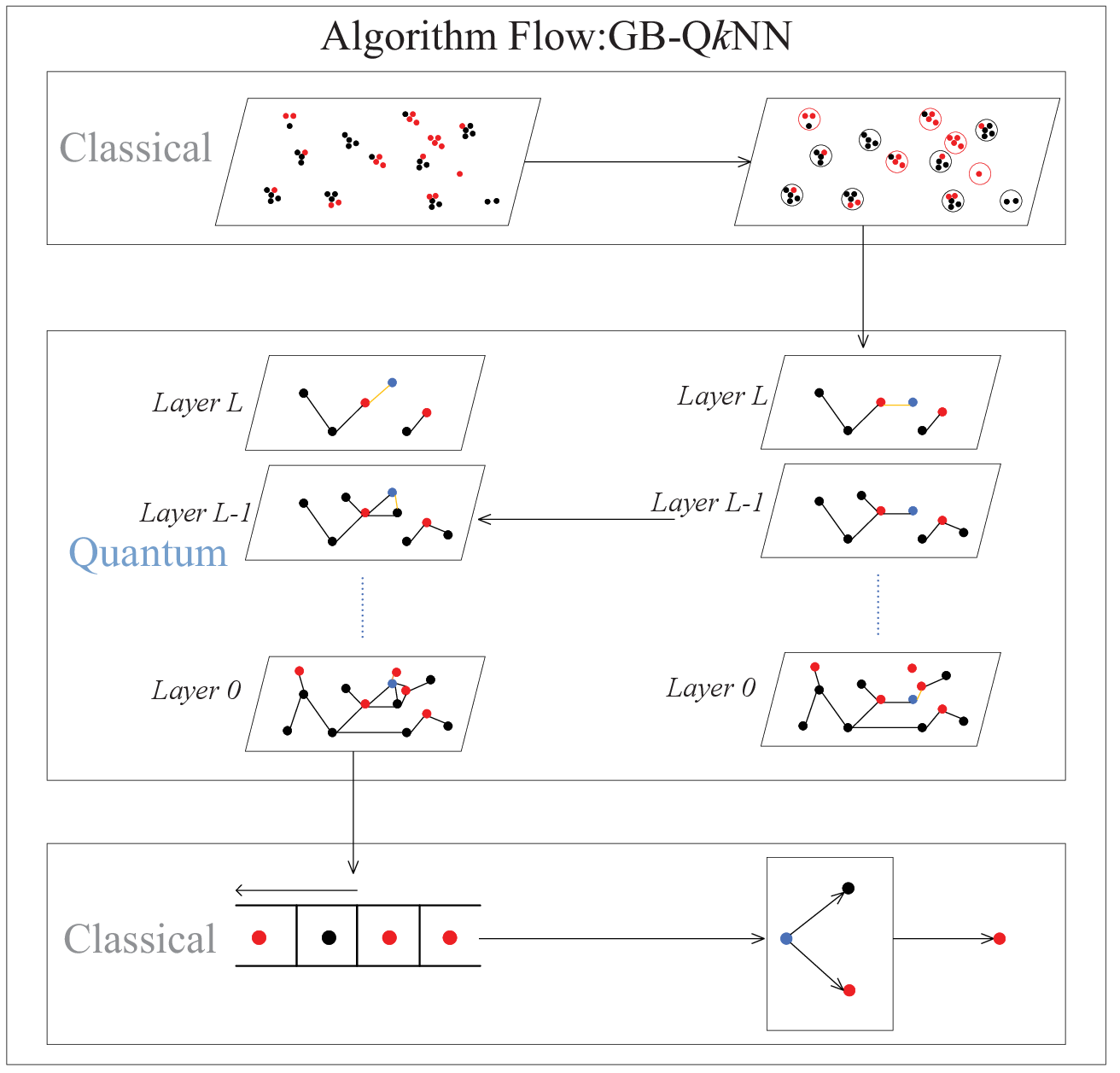}
	\caption{\label{fig_6} Overall flow of the GB-Q$k$NN.}
\end{figure}

The overall algorithm flow of the proposed GB-Q$k$NN algorithm is depicted by the flow chart as shown in Fig. \ref{fig_6}. 

The process is divided into three main stages. The first stage involves the classical algorithm: the original data set $D$ is used to generate the granular ball, and the granular ball data set $G$ is obtained. The second stage is based on the quantum algorithm, which comprises two key steps: constructing the quantum HNSW and searching within the quantum HNSW. These steps use quantum algorithms such as QRAM, angle encoding, Swap test, and quantum comparators to facilitate parallel distance calculations and magnitude comparisons during the construction of the HNSW for the granular-ball data set $G$. Subsequently, these quantum algorithms are employed to compute the distance between the test point and the original points in the graph and identify the neighboring points of the test point, thereby completing the quantum HNSW search process. The third stage involves classical computation to determine the type distribution within the priority queue $Q$. The test point is classified according to the type with the highest proportion in the priority queue. The following sections provide a detailed explanation of each stage.
The process is divided into three main stages. The first stage involves the classical algorithm: the original data set $D$ is used to generate the granular ball, and the granular ball data set $G$ is obtained. The second stage is based on the quantum algorithm, which comprises two key steps: constructing the quantum HNSW and searching within the quantum HNSW. These steps use quantum algorithms such as QRAM, angle encoding, Swap test, and quantum comparators to facilitate parallel distance calculations and magnitude comparisons during the construction of the HNSW for the granular-ball data set $G$. Subsequently, these quantum algorithms are employed to compute the distance between the test point and the original points in the graph and identify the neighboring points of the test point, thereby completing the quantum HNSW search process. The third stage involves classical computation to determine the type distribution within the priority queue $Q$. The test point is classified according to the type with the highest proportion in the priority queue. The following sections provide a detailed explanation of each stage.

\subsection{Granular-ball Generation}
The original data set $D$ is initially considered to be the first granular-ball. A random point is selected from the granular-ball to serve as the center, and another point, which is the most distant and belongs to a different class, is chosen as the second center. The purity of each granular-ball is then assessed to determine whether it meets the threshold value $T$. If the purity is insufficient, the granular-ball is split according to the following rule: the remaining data points within the current granular-ball are assigned to the granular-ball corresponding to the center that is closest to them. For the newly created granular-ball resulting from the split, the original center is retained as the new center, and another point, the farthest from the current center and belonging to a different class, is selected as the second center. This splitting process continues iteratively until all granular-ball exhibit a purity greater than or equal to the threshold $T$. The final data set of granular-balls $G$ is obtained, with its size denoted as $M$.

\subsection{Quantum HNSW Construction}  

Quantum HNSW construction is the basis of quantum HNSW search. We need to obtain the data sets ${S_i}(i \in \left[ {0,\log M} \right]$ \cite{malkov-hnsw} of each layer, which represents the $i$th layer), and the adjacency matrix ${J_i}(i \in \left[ {0,\log M} \right])$ of each layer from the quantum HNSW construction.

\begin{figure}[!t]
	\centering	\includegraphics[width=0.5\linewidth]{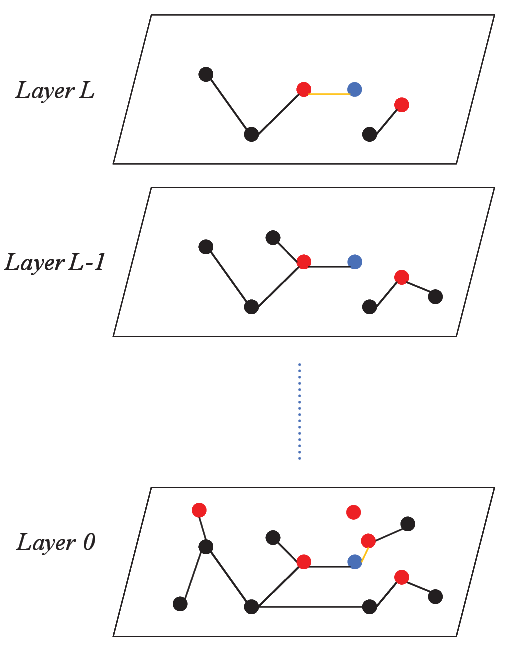}
	\caption{\label{fig_5} HNSW method.}
\end{figure}

Figure \ref{fig_5} shows the process of one data point participating in the construction of quantum HNSW, where the blue point is the newly inserted data point. The specific construction process of one data point is described as follows.  

\paragraph{Step 1: Determine the top layer to insert.} The top layer of insertion ${L}$ is first determined by a random function as in Eq. \ref{l_temp}.
\begin{equation}
    \label{l_temp}
    L_{temp} = \left\lfloor - {{\log }_2}(r) \right\rfloor
\end{equation}
where $r \in (0,1)$ is a uniformly distributed random number. To ensure that ${L}$ does not exceed the total number of layers, the final definition of L is given in Eq. \ref{L}.
\begin{equation}
    \label{L}
    {L} = \min (L_{temp},\log M)
\end{equation}

\paragraph{Step 2: Insert one data point from layer to layer.} The data point will construct the graph from the layer ${L}$ down. In the construction process, only the insertion operation is needed if there is only the newly inserted data point in the layer. If there are other data points in the layer, the similarity calculation will determine the nearest data point to the insertion point. In addition, the connection relationship between the insertion point and the nearest data point is established, and the nearest data point is used as the neighbor node of the insertion point in the next layer. When the insertion point is inserted into the next layer, the nearest data point is determined by the neighboring nodes obtained from the previous layer and the adjacency matrix of the current layer. The insertion point establishes a connection relationship with the nearest data point through the adjacency matrix. Similarly, the nearest point will be the neighbor node of the next layer. This operation continues until the insertion point is inserted into the lowest layer. 

When all data points in $G$ are inserted through the two steps above, the data set ${S_i}$ and the adjacency matrix ${J_i}$ of each layer are obtained.

The pseudocode for the quantum HNSW graph construction is shown in Algorithm 3.

\begin{algorithm}[!t]
    \scriptsize
	\caption{Quantum HNSW Construction}
	\label{alg:qhnsw_construction}
	\textbf{Input}: The size of Granular-ball data set $M$, granular-ball data set $G$, maximum neighbor count $m$\\
	\textbf{Output}: Data point set $S[i][j]$ $(i \in [0, \log M])$, adjacency matrix $J[i][j][k]$ $(i \in [0, \log M])(j,k \in [0, M])$
	\begin{algorithmic}[0]
		\STATE Initialize $S \leftarrow \emptyset$, $J \leftarrow \emptyset$, $count_p \leftarrow 0$
		\WHILE{$count_p < M$}
		\STATE Sample $r \sim \text{Uniform}(0, 1)$
		\STATE Compute $L \leftarrow \min(-\log_2(r), \log M)$
		\STATE Set $layer \leftarrow 0$
		\WHILE{$layer < L$}
		\STATE Add $G[count_p]$ to $S[layer]$
		\STATE Initialize $min \leftarrow \text{INT\_MAX}$, $L_c[layer] \leftarrow \emptyset$, $D_c[count_p] \leftarrow \emptyset$
		\STATE Initialize$H\leftarrow\text{Algorithm}\ref{alg:swaptest}(S[layer], G[count_p])$
		\STATE Set $temp_m \leftarrow m$
		\WHILE{$temp_m > 0$}
		\IF{$H.\text{size}() \neq 0$}
		\IF{$min \neq \text{Algorithm}\ref{alg:QuantumCompare}(\text{min}, H.\text{front}())$}
		\STATE $min \leftarrow H.\text{front}()$
		\STATE $index \leftarrow H.\text{front().index()}$
		\ENDIF
		\STATE Remove $H.\text{front}()$ from $H$
		\ELSE 
		\STATE Break;
		\ENDIF
		\STATE $temp_m \leftarrow temp_m - 1$
		\ENDWHILE
		\STATE $J[layer][count_p][\text{index}] \leftarrow 1$
		\STATE $layer \leftarrow layer + 1$
		\ENDWHILE
		\STATE $count_p \leftarrow count_p + 1$
		\ENDWHILE
		\STATE \textbf{return} $S, J$
	\end{algorithmic}
\end{algorithm}

After introducing the whole construction process from the perspective of how each point is inserted, we will introduce the construction process of each layer from the perspective of how it is implemented through quantum algorithms. Firstly, the data are encoded by QRAM, angle encoding. The similarity measurement is performed on the encoded data, and the connection relationship is established on the basis of the similarity. Then, this data point is used as the next layer's neighbor node for the insertion point. The steps are described below.

\begin{figure}[!t]
	\centering	\includegraphics[width=0.95\linewidth]{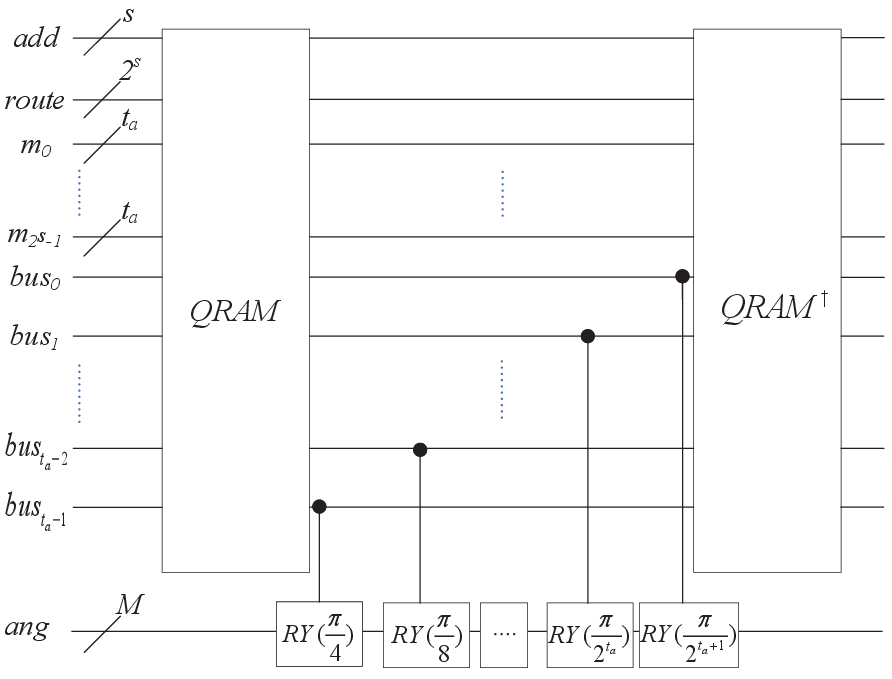}
	\caption{\label{fig_7} Quantum state encoding circuit.}
\end{figure}

\paragraph{Step1: Quantum state encoding.} Suppose that the dimension of the granular-ball data set $G$ is $d$, and the range of data point values for each dimension is $\left\lfloor 0, {2^{{t_a}}-1} \right\rfloor$. Firstly, the granular-ball data set $G$ is stored in the memory qubits by QRAM, and the address qubits are used to control the memory qubits in parallel. Subsequently, the angle encoding is used to map the data on the bus qubits to the angle on the Bloch sphere. The specific quantum circuit is shown in Fig. \ref{fig_7}. From top to bottom, the first $s$ qubits store the address information of one dimension, where ${2^s} \ge M$. The address qubits can be reused for $d$ dimensions of the same data. The next $2^s$ route qubits are used as a bridge for the address qubits to find the memory qubits. There are $2^s$ groups of memory qubits and every group of ${t_a}$ memory qubits is used to store specific numerical information. The following ${t_a}$ bus qubits are adopted to store numerical information, which will be used in the subsequent angle encoding. Finally, qubits $ang$ are added to store the angle encoding information.

It can be seen from Fig. \ref{fig_7} that the data set $G$ is first encoded by QRAM, whose specific quantum circuit is shown in \cite{giovannetti-qram}. This is followed by angle encoding. The data is stored in binary form on the bus qubits of QRAM, which can be encoded into angles by controlled rotation operations. When the control qubit is 1, the controlled rotation operations can encode the data into angles $\frac{\pi }{{{2^{v}}}}$ of the qubit $ang$. The value of $v$ starts from 2 and increases by 1 each time until $t_a+1$. Finally, we use $QRA{M^\dag }$ to restore the quantum state to the state before using QRAM.

\paragraph{Step2: Obtaining similarity.} The similarity of the two quantum states is their inner product, which can be obtained by performing a Swap test operation.  

First, we should obtain the quantum states that will participate in the similarity calculation. They are the node set information in each layer and the newly inserted data point, which are encoded in qubits $ang$ in Fig. \ref{fig_7}.
Based on the data sets ${S_i}$, the adjacency matrix ${J_i}$ and the neighboring nodes of each layer, we can obtain the set of nodes ${S_{ci}}(c \in [0, m], i \in [0, \log M])$. The node set of the current layer is the union of the neighbor node and the one-hop nodes of the neighbor node. 

\begin{figure}[!t]
	\centering	\includegraphics[width=0.7\linewidth]{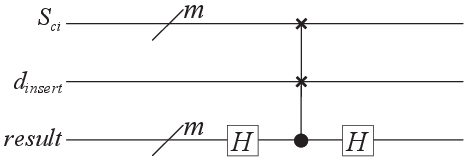}
	\caption{\label{fig_8} Similarity computing quantum circuit.}
\end{figure}

Then, as shown in Fig. \ref{fig_8}, the angle qubits corresponding to the current insertion point ${d_{insert}}$ and the set of points ${S_{ci}}$ are used as input for the similarity calculation. Each angle qubit in the set ${S_{ci}}$ performs the Swap test with ${d_{insert}}$. Information about their similarity ${\left| {\left\langle {{{d_{insert}}}}\mathrel{\left | {\vphantom {{{d_{insert}}} {{S_{ci}}}}} \right. \kern-\nulldelimiterspace}
{{{S_{ci}}}} \right\rangle } \right|^2}$ 
is stored in the probability of measuring qubits $result$ to get 0 or 1.

The probability expression related to the similarity in Swap test is specific as follows:
\begin{equation}
	P(0) = \frac{1}{2} + \frac{1}{2}{\left| {\left\langle {{{d_{insert}}}}
 \mathrel{\left | {\vphantom {{{d_{insert}}} {{S_{ci}}}}}
 \right. \kern-\nulldelimiterspace} {{{S_{ci}}}} \right\rangle } \right|^2}
\end{equation}
\begin{equation}
	P(1) = \frac{1}{2} - \frac{1}{2}{\left| {\left\langle {{{d_{insert}}}}
 \mathrel{\left | {\vphantom {{{d_{insert}}} {{S_{ci}}}}}
 \right. \kern-\nulldelimiterspace} {{{S_{ci}}}} \right\rangle } \right|^2}
\end{equation}
where ${\left| {\left\langle {{{d_{insert}}}}\mathrel{\left | {\vphantom {{{d_{insert}}} {{S_{ci}}}}} \right. \kern-\nulldelimiterspace}
{{{S_{ci}}}} \right\rangle } \right|^2}$ represents the similarity between ${S_{ci}}$ and ${d_{insert}}$. $p(0)$ and $p(1)$ represent the probability of obtaining the state $\vert{0}\rangle$ and $\vert{1}\rangle$ after quantum measurement. 

The lower the probability that we get $result$ to be 1 when we measure, the more similar the two data points are and the closer they are to each other.

\paragraph{Step 3: Building connections.} We must find the data point $C$ closest to the insertion point ${d_{insert}}$ in the current layer. $C$ can be obtained by using a quantum comparator \cite{yuan-quantum-image-segmentation} on the similarity information obtained in Step 2. Then, the corresponding element is set to 1 in the adjacency matrix ${J_i}$ to represent the connection relationship between the two points. Finally, the point $C$ is used as the neighbor node inserted in the next layer.

Once all points in the granular-ball data set have been inserted, the construction of the quantum HNSW is complete, resulting in the constructed data sets ${S_i}$ of each layer and the adjacency matrix ${J_i}$ for each layer.

\subsection{Quantum HNSW Search}
The search will begin from the top layer of the constructed quantum HNSW for the test data point ${d_t}$ based on ${S_i}$ and ${J_i}$ previously determined. The pseudocode of the quantum HNSW search is shown in Algorithm 4. 

We divide the quantum HNSW search process into four steps. 

\paragraph{Step1: Quantum state encoding.} The data in the constructed graph has been encoded into angle qubits during the quantum HNSW construction. We only need to encode the test data point ${d_t}$ in a quantum state using QRAM and angle encoding, in the same way as in Fig. \ref{fig_7}.

\begin{figure}[!t]
 	\centering	\includegraphics[width=0.70\linewidth]{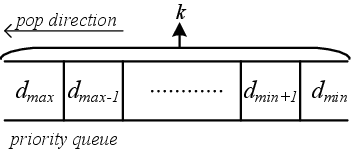}
	\caption{\label{fig_10} Priority queue structure.}
\end{figure}

\paragraph{Step2: Obtain the nearest data point of the current layer.} To determine the nearest data point of ${d_t}$ in the current layer, we need to obtain the node set ${S_{ci}}(c \in \left[ {0,m} \right], i \in \left[ {0,\log M} \right])$ to perform similarity with ${d_t}$. Based on data sets ${S_i}$ and the adjacency matrix ${J_i}$ established during the construction process for each layer, we can obtain the nearest data point of the test point ${d_t}$ in the current layer using the same method as in the construction process. Then, this data point will be added to a predefined priority queue $Q$. The size of the queue is $k$, and the priority queue will automatically be sorted by the values of members, and the higher values have a higher priority. The specific structure of the queue is shown in Fig. \ref{fig_10}. 
\begin{algorithm}[!t]
    \small
	\caption{Quantum HNSW Search}
	\label{alg:qhnswsearch}
	\textbf{Input}: Data point set $S[i][j]$ $(i \in 0 \text{ to } \log M)$, adjacency matrix $J[i][j][k]$ $(i \in 0 \text{ to } \log M)$, test point $d_t$, granular-ball data set size $M$, granular-ball data set $G$, ma3ximum neighbor count $m$\\
	\textbf{Output}: Type of test point $d_t$
	\begin{algorithmic}[0]
		\STATE Initialize $H \gets \emptyset$ 
		\STATE $layer \gets 0$ 
		\STATE $neighborpos \gets 0$ 
		\STATE $Q_\text{priority} \gets \emptyset$
		\STATE $typetest \gets \emptyset$
		\WHILE{$layer < \log M$}
		\STATE Initialize $similarityCalcSet \gets \emptyset$, $min \leftarrow \text{INT\_MAX}$
		\IF{$layer == 0$} 
		\STATE $H \gets \text{Algorithm}\ref{alg:swaptest}(S[layer], d_t)$
		\ELSE
		\STATE $index \gets 0$
		\STATE $count \gets 0$
		\WHILE{$index < M$ \AND $count < m$}
		\IF{$J[layer][index][neighborpos] == 1$}
		\STATE Add $G[index]$ to $similarityCalcSet$
		\STATE $count \gets count + 1$
		\ENDIF
		\STATE $index \gets index + 1$
		\ENDWHILE
		\STATE $H \gets \text{Algorithm}\ref{alg:swaptest}(similarityCalcSet, d_t)$
		\ENDIF
		\STATE $tempm \gets m$ 
		\WHILE{$tempm > 0$}
		\STATE $tempm \gets tempm - 1$
		\IF{$H.\text{size()} \neq 0$}
		\IF{$min \neq \text{Algorithm}\ref{alg:QuantumCompare}(min, H.\text{front()})$}
		\STATE $min \gets H.\text{front()}$
		\STATE $index \gets H.\text{front()}.\text{index()}$ 
		\ENDIF
		\STATE Remove $H.\text{front()}$ from $H$
		\ELSE 
		\STATE Break;
		\ENDIF
		\ENDWHILE
		\STATE $neighborpos \gets index$ 
		\IF{$Q_\text{priority}.\text{size()} > m$ \AND $min <= Q_\text{priority}.\text{front()}$}
		\STATE $Q_\text{priority}.\text{popfront()}$
		\ENDIF
		\STATE Add $G[neighborpos]$ to $Q_\text{priority}$
		\STATE $layer \gets layer + 1$
		\ENDWHILE
		\STATE Compute $type \gets max_\text{type}(Q_\text{priority}) $
		\STATE \textbf{return} $type$
	\end{algorithmic}
\end{algorithm} 

\paragraph{Step3: Obtain $k$ nearest data points.} We traverse the quantum HNSW graph from the top to the bottom layer to obtain the $k$ nearest neighbors of the test data point ${d_t}$ by repeatedly executing Step 1 and Step 2. The nearest data point is added to the priority queue $Q$ for each layer. When the capacity of $Q$ exceeds $k$, a quantum comparator is used to evaluate the relationship between the nearest data point ${d_{ci}}(i \in (0,\log M))$ and the highest priority data ${d_{bi}}(i \in (0,\log M))$ in the queue. If ${d_{ci}} > {d_{bi}}$, the priority queue remains unchanged; otherwise, ${d_{ci}}$ is added to the queue and ${d_{bi}}$ is removed. 

\paragraph{Step4: Classification of test data.} Based on the $k$ nearest neighbors obtained in Step 3, the label of the test data point ${d_t}$ is determined by the majority class label of its nearest $k$ neighbors.

\section{Time Complexity Analysis}
This section analyzes the time complexity in two aspects: construction and search. 
\subsection{Construction Part}
The algorithm begins by generating a granular-ball data set from the classical data set, and the time complexity to generate the classical granular ball is given by $O(cdN)$ \cite{xia-granular-ball}, which is performed only once as part of the data pre-processing phase. Then, we will analyze the construction complexity of the quantum HNSW based on the granular-ball. First, the time complexity of QRAM, based on its quantum circuit structure and characteristics, is $O(\log M)$. Second, the time complexity of angle encoding, which involves only rotation operations, is $O(1)$. Third, the time complexity of the Swap test is $O(1)$, and the time complexity of the quantum iterative comparison is $O(m)$, where $m$ represents the maximum number of neighbors for each node, which is much smaller than $M$.

The loop process consists of two parts. The first part involves all $M$ points participating in the construction. The second part requires that each point begin construction from its corresponding ${L}$ ($L \le \log M$) layer, with operations including angle encoding, Swap test, and iterative comparison. Based on the above analysis, the time complexity of the loop part is $O(M)O(\log M)(O(m) + O(1) + O(1))$, which simplifies to $O(M)O(\log M)$. Including QRAM and granular-ball generation, the resulting time complexity is $O(M)O(\log M) + O(\log M) + O(cdN)$, which can be simplified to $O(cdN)$. 


\subsection{Search Part}
 The input test point begins at the top layer and searches a total of $\log M$ layers to the bottom. Each layer involves angle encoding, Swap test, and quantum iterative comparators. The time complexity of each quantum algorithm is shown in the construction part. Therefore, the time complexity of the loop search process is $O(\log M)(O(m) + O(1) + O(1))$. Since the maximum number of neighbors $m$ is much smaller than the size of the granular-ball data set $M$, this simplifies to $O(\log M)$. Thus, the final time complexity for the quantum HNSW search is $O(\log M)$. 
\subsection{Time Complexity Comparison}
To evaluate the performance of the GB-Q$k$NN algorithm, the comparison between the proposed algorithm and the other quantum $k$NN algorithms is performed, which is shown in Table \ref{quantum}. $M$ denotes the size of the granular-ball data set, and $N$ represents the size of the original data set, where $N \gg M$. It can be seen from Table \ref{quantum} that the proposed GB-Q$k$NN algorithm exhibits the lowest time complexity. 

\begin{table}[!t]
	\centering
	
	\label{table 2}
	\begin{tabular}{ll}
		\toprule
		Methods    & Time complexity           \\
		\midrule
		FQ$k$NN \cite{basheer-quantum-knn}    & $O\sqrt{kN})$          \\
		SQ$k$NN \cite{quezada-qknn-sorting}    & $O(kN)$                \\
		PQ$k$NN \cite{feng-polar-distance}    & $O(\sqrt{kN})$         \\
		MQ$k$NN \cite{gao-mahalanobis-knn}    & $O(\sqrt{kN}+\log{N})$ \\
		EQ$k$NN \cite{zardini-euclidean-knn}    & $O(N+\log{N})$         \\
		GB-Q$k$NN &$O(\log{M})$            \\
		\bottomrule
	\end{tabular}
        \caption{\label{quantum}Comparisons of the time complexity between the proposed GB-Q$k$NN algorithm and the other quantum $k$NN algorithms}
\end{table}

Table \ref{class} compares the proposed algorithm with the classical graph-based approximate $k$NN algorithms. It shows that the GB-Q$k$ NN algorithm outperforms the current classical graph-based approximate $k$NN algorithms in construction and search complexity. This highlights the superior theoretical performance and faster execution speed of the proposed algorithm.
\begin{table}[!t]
	\tiny 
	\centering
	
	\label{table 3}
	\begin{tabular}{p{3.4cm}p{2.0cm}p{1.63cm}}
		\toprule
		Methods & Construct Complexity & Search Complexity\\
		\midrule
		DPG \cite{li-approximate-knn}  & $O({N^{1.14}} + N)$ & $O({N^{0.28}})$\\
		HCNNG \cite{munoz-hierarchical-graphs} & $O(kN)$ &  $O({N^{0.4}})$\\
		Vamana \cite{jayaram-diskann} &$O({N^{1.16}})$ & $O({N^{0.75}})$\\
		NSSG \cite{fu-satellite-graph} &$O({N^{1.14}} + N)$ & $O(\log N)$\\
		HNSW \cite{malkov-hnsw} &$O(N\log N)$ & $O(\log N)$\\
		GB-Q$k$NN &$O(cdN)$ & $O(\log M)$\\
		\bottomrule
	\end{tabular}
        \caption{\label{class}Comparisons of the time complexity between the proposed GB-Q$k$NN algorithm and classical $k$NN algorithms}
\end{table}

\section{Conclusion and Future Works}
This paper proposes a GB-Q$k$NN algorithm to improve the search efficiency of $k$NN in large-scale data sets. This approach integrates granular-ball theory, the HNSW method, and quantum computing to reduce the time complexity and improve the search efficiency of $k$NN. The proposed GB-Q$k$NN algorithm has a very low time complexity in the search process. Future work will focus on optimizing the granular-ball generation and attribute-reduction processes to reduce time complexity. With the ongoing advancement of quantum computing, the GB-Q$k$ NN algorithm has promising application potential.

\section*{Acknowledgments}

The authors would like to acknowledge the financial support of the National Natural Science Foundation of China (Nos. 62221005, 62450043, 62222601, and 62176033).

\bibliographystyle{named}
\bibliography{ijcai25}

\begin{thebibliography}{}

\bibitem[\protect\citeauthoryear{Altaisky}{2001}]{altaisky-qnn}
M.~V. Altaisky.
\newblock Quantum neural network.
\newblock {\em arXiv preprint}, quant-ph/0107012, 2001.

\bibitem[\protect\citeauthoryear{Basheer \bgroup \em et al.\egroup }{2020}]{basheer-quantum-knn}
A.~Basheer, S.~K. Goyal, and A.~Afham.
\newblock Quantum $k$-nearest neighbors algorithm.
\newblock {\em arXiv preprint}, arXiv:2003.09187, 2020.

\bibitem[\protect\citeauthoryear{Bennett \bgroup \em et al.\egroup }{1993}]{bennett-teleportation}
C.~H. Bennett, G.~Brassard, and C.~Crépeau.
\newblock Teleporting an unknown quantum state via dual classical and einstein-podolsky-rosen channels.
\newblock {\em Physical Review Letters}, 70(13):1895, 1993.

\bibitem[\protect\citeauthoryear{Farhi \bgroup \em et al.\egroup }{2017}]{farhi-fixed-architectures}
E.~Farhi, J.~Goldstone, and S.~Gutmann.
\newblock Quantum algorithms for fixed qubit architectures.
\newblock {\em arXiv preprint}, arXiv:1703.06199, 2017.

\bibitem[\protect\citeauthoryear{Feng \bgroup \em et al.\egroup }{2023}]{feng-polar-distance}
C.~Feng, B.~Zhao, and X.~Zhou.
\newblock An enhanced quantum k-nearest neighbor classification algorithm based on polar distance.
\newblock {\em Entropy}, 25(1):127, 2023.

\bibitem[\protect\citeauthoryear{Fu \bgroup \em et al.\egroup }{2021}]{fu-satellite-graph}
C.~Fu, C.~Wang, and D.~Cai.
\newblock High dimensional similarity search with satellite system graph: Efficiency, scalability, and unindexed query compatibility.
\newblock {\em IEEE Transactions on Pattern Analysis and Machine Intelligence}, 44(8):4139--4150, 2021.

\bibitem[\protect\citeauthoryear{Gao \bgroup \em et al.\egroup }{2022}]{gao-mahalanobis-knn}
L.~Z. Gao, C.~Y. Lu, and G.~D. Guo.
\newblock Quantum k-nearest neighbors classification algorithm based on mahalanobis distance.
\newblock {\em Frontiers in Physics}, 10:1047466, 2022.

\bibitem[\protect\citeauthoryear{Giovannetti \bgroup \em et al.\egroup }{2008}]{giovannetti-qram}
V.~Giovannetti, S.~Lloyd, and L.~Maccone.
\newblock Quantum random access memory.
\newblock {\em Physical Review Letters}, 100(16):160501, 2008.

\bibitem[\protect\citeauthoryear{Guo \bgroup \em et al.\egroup }{2003}]{guo-knn-model}
G.~Guo, D.~Bell, and H.~Wang.
\newblock Knn model-based approach in classification.
\newblock In {\em OTM Confederated International Conferences "On the Move to Meaningful Internet Systems"}. Springer, Berlin, Heidelberg, 2003.

\bibitem[\protect\citeauthoryear{Jayaram Subramanya~S.}{2019}]{jayaram-diskann}
Simhadri H.~V. Jayaram Subramanya~S., Devvrit~F.
\newblock Diskann: Fast accurate billion-point nearest neighbor search on a single node.
\newblock In {\em Advances in Neural Information Processing Systems}, volume~32, 2019.

\bibitem[\protect\citeauthoryear{Li \bgroup \em et al.\egroup }{2022}]{li-hamming-knn}
J.~Li, S.~Lin, and K.~Yu.
\newblock Quantum k-nearest neighbor classification algorithm based on hamming distance.
\newblock {\em Quantum Information Processing}, 21(1):18, 2022.

\bibitem[\protect\citeauthoryear{Li~W. and others.}{2019}]{li-approximate-knn}
Sun~Y. Li~W., Zhang~Y. and others.
\newblock Approximate nearest neighbor search on high dimensional data—experiments, analyses, and improvement.
\newblock {\em IEEE Transactions on Knowledge and Data Engineering}, 32(8):1475--1488, 2019.

\bibitem[\protect\citeauthoryear{Malkov and Yashunin}{2016}]{malkov-hnsw}
Y.~A. Malkov and D.~A. Yashunin.
\newblock Efficient and robust approximate nearest neighbor search using hierarchical navigable small world graphs.
\newblock {\em IEEE Transactions on Pattern Analysis and Machine Intelligence}, PP(99), 2016.

\bibitem[\protect\citeauthoryear{Munoz J.~V.}{2019}]{munoz-hierarchical-graphs}
Dias~Z. Munoz J.~V., Gonçalves M.~A.
\newblock Hierarchical clustering-based graphs for large scale approximate nearest neighbor search.
\newblock {\em Pattern Recognition}, 96:106970, 2019.

\bibitem[\protect\citeauthoryear{Quezada \bgroup \em et al.\egroup }{2022}]{quezada-qknn-sorting}
L.~F. Quezada, G.~H. Sun, and S.~H. Dong.
\newblock Quantum version of the $k$‐nn classifier based on a quantum sorting algorithm.
\newblock {\em Annalen der Physik}, 534(5):2100449, 2022.

\bibitem[\protect\citeauthoryear{Rebentrost \bgroup \em et al.\egroup }{2014}]{rebentrost-qsvm}
P.~Rebentrost, M.~Mohseni, and S.~Lloyd.
\newblock Quantum support vector machine for big data classification.
\newblock {\em Physical Review Letters}, 113(13):130503, 2014.

\bibitem[\protect\citeauthoryear{Xia \bgroup \em et al.\egroup }{2022}]{xia-granular-ball}
S.~Xia, X.~Dai, and G.~Wang.
\newblock An efficient and adaptive granular-ball generation method in classification problem.
\newblock {\em IEEE Transactions on Neural Networks and Learning Systems}, 35(4):5319--5331, 2022.

\bibitem[\protect\citeauthoryear{Yuan \bgroup \em et al.\egroup }{2022}]{yuan-quantum-image-segmentation}
S.~Yuan, W.~Zhao, S.~Gao, et~al.
\newblock An adaptive threshold-based quantum image segmentation algorithm and its simulation.
\newblock {\em Quantum Information Processing}, 21(10):359, 2022.

\bibitem[\protect\citeauthoryear{Zardini \bgroup \em et al.\egroup }{2024}]{zardini-euclidean-knn}
E.~Zardini, E.~Blanzieri, and D.~Pastorello.
\newblock A quantum $k$-nearest neighbors algorithm based on the euclidean distance estimation.
\newblock {\em Quantum Machine Intelligence}, 6(1):1--22, 2024.

\end{thebibliography}

\end{document}